\documentstyle[12pt]{article}
\def\v{\begingroup\obeyspaces\u}
\def\u#1{\verb!#1!\endgroup}

\def\IP{\v{IPROC}}

\def\HW{{\small HERWIG}}

\def\JS{{\small JETSET}}
\def\MS{{\small MSSM}}
\def\SY{{\small SUSY}}

\def\TA{{\small TAUOLA}}

\begin{document}
\tolerance=100000
\thispagestyle{empty}
\setcounter{page}{0}
 \begin{flushright}
Cavendish-HEP-01/16\\
CERN-TH/2001-369\\
DAMTP-2001-109\\
IPPP/01/65\\
KEK--TH-797\\
MPI-PhT/2002-02\\
December 2001
\end{flushright}

\begin{center}
{\Large \bf HERWIG 6.4 Release Note}\\[4mm]

{G. Corcella\\[0.4mm]
\it Max Planck Institute f\"{u}r Physik, Munich\\[0.4mm]
E-mail: \tt{corcella@mppmu.mpg.de}}\\[4mm]

{I.G.\ Knowles\\[0.4mm]
\it Department of Physics and Astronomy, University of Edinburgh\\[0.4mm]
E-mail: \tt{iknowles@supanet.com}}\\[4mm]

{G.\ Marchesini\\[0.4mm]
\it Dipartimento di Fisica, Universit\`a di Milano-Bicocca, and I.N.F.N.,
Sezione di Milano\\[0.4mm]
E-mail: \tt{Giuseppe.Marchesini@mi.infn.it}}\\[4mm]

{S.\ Moretti\\[0.4mm]
\it Theory Division, CERN, and Institute for Particle Physics Phenomenology,
University of Durham\\[0.4mm]
E-mail: \tt{stefano.moretti@cern.ch}}\\[4mm]

{K.\ Odagiri\\[0.4mm]
\it Theory Division, KEK\\[0.4mm]
E-mail: \tt{odagirik@post.kek.jp}}\\[4mm]

{P.\ Richardson\\[0.4mm]
\it Department of Applied Mathematics and Theoretical Physics and\\
Cavendish Laboratory, University of Cambridge\\[0.4mm]
E-mail: \tt{richardn@hep.phy.cam.ac.uk}}\\[4mm]

{M.H.\ Seymour\\[0.4mm]
\it Department of Physics and Astronomy,
University of Manchester\\[0.4mm]
E-mail: \tt{M.Seymour@rl.ac.uk}}\\[4mm]

{B.R.\ Webber\\[0.4mm]
\it Cavendish Laboratory, University of Cambridge\\[0.4mm]
E-mail: \tt{webber@hep.phy.cam.ac.uk}}\\[4mm]

\end{center}

\vspace*{\fill}

\begin{abstract}{\small\noindent
    A new release of the Monte Carlo program \HW\ (version 6.4) is now
    available. The main  new  features  are: spin  correlations  between the
    production and decay of heavy  fermions, i.e.\ top quarks, $\tau$
    leptons and \SY\ particles;  polarization effects in \SY\ production
    processes in lepton-lepton  collisions; an interface to \TA\ for $\tau$
    decays; \MS\ Higgs processes in lepton-lepton collisions.}
\end{abstract}

\vspace*{\fill}
\newpage
\tableofcontents
\setcounter{page}{1}

\section{Introduction}

The last major public version (6.2) of \HW\ was reported in detail
in \cite{AllHW}.  The new features of version 6.3 are described
in \cite{Corcella:2001pi}.  In this note we describe the main modifications
and new features included in the latest public version, 6.4.

Please refer to \cite{AllHW} and to the present paper if
using version 6.4 of the program.
When running \MS\ processes starting from version 6.1, please add 
reference to \cite{SUSYpap}.

\subsection{Availability}
The new program, together  with other useful files and information,
can be obtained from the following web site:
\small\begin{quote}\tt
            http://hepwww.rl.ac.uk/theory/seymour/herwig/
\end{quote}\normalsize
    This will temporarily be mirrored at CERN for the next few weeks:
\small\begin{quote}\tt
            http://home.cern.ch/seymour/herwig/
\end{quote}\normalsize

\section{Spin correlations}

    Spin  correlation  effects  have been  added in  processes where top
    quarks, $\tau$ leptons and \SY\ particles are produced, as described in
    \cite{Richardson:2001df}. At the moment the effects  are only calculated
    for the production of these particles  in the  following  processes:
    \IP=100-199,  \IP=700-799,   \IP=1300-1399,   \IP=1400-1499,
    \IP=1500-1599,\IP=1700-1799, \IP=2000-2099,   \IP=2800-2825,
    \IP=3000-3030. However, if these particles  are  produced in other
    processes, the  spin  correlation  algorithm will still  be  used to
    perform  their  decays. The correlations  are also  included for the
    decay of the \MS\ Higgs bosons, regardless of how they are produced.

    The spin correlations are controlled by the logical
    variable\footnote{Default values for input variables are
      shown in square brackets.} \v{SYSPIN} [\v{.TRUE.}]
    which  switches  the  correlations  on. If  required the 
    correlations are initialized by the new routine \v{HWISPN}. This routine
    initializes the two, three and four body matrix elements.

    The three and four  body matrix  elements can be  used separately to
    generate  the decay  distributions without spin correlation effects.
    These are switched on by the switches \v{THREEB} [\v{.TRUE.}]
for three body
    decay and \v{FOURB} [\v{.FALSE.}] for four body decays.
The four body decays
    are only important  in \SY\ Higgs studies,  and have small branching
    ratios. However, they take some time to initialize and are therefore
    switched off by default.

   The  initialization  of the spin  correlations  and/or  decay matrix
    elements  can be  time  consuming  and we have therefore included an
    option to read/write the information.  The information is written to
    unit \v{LWDEC} [88] and read from \v{LRDEC} [0]. If either are zero
the data
    is not  written/read.  If \v{IPRINT}=2 then information on the branching
    ratios  for the decay modes  and the maximum  weights for the matrix
    elements is outputted.

{\bf Important note:} If the spin correlation (\v{SYSPIN}) or matrix element 
switches  (\v{THREEB}, \v{FOURB}) are \v{.TRUE.}, then the matrix element
codes (\v{NME} entries) for the decays concerned are not used;
the calculated matrix elements are used instead.

For top decays via the weak matrix element, when \v{SYSPIN}=\v{.TRUE.} the spin
correlation algorithm uses the helicity amplitudes to perform the decay,
whereas when \v{SYSPIN}=\v{.FALSE.} the spin averaged matrix element is used.

Top decay to either a real or virtual Higgs boson is not currently
implemented in the spin correlation
algorithm and therefore the spin correlations should be switched off,
\v{SYSPIN}=\v{.FALSE.}, if you wish to study this decay process.

\subsection{Polarized lepton beams}
    The effect of polarization for incoming leptonic beams in
\MS\ \SY\ processes has also been included. These effects are included both
    in the  production  of \SY\  particles  and via  the  spin  correlation
    algorithm in their decays.

\section{New MSSM Higgs processes}
    The  following \MS\ Higgs  production  processes in $\ell^+\ell^-$
 collisions  have been added ($\ell=e,\mu$):

\begin{table}[h!]
\begin{center}
\begin{tabular}{|c|l|}
\hline
\v{IPROC} & Process\\
\hline
    910    &      $\ell^+ \ell^- \to \nu_e \bar\nu_e h^0 + e^+ e^- h^0$\\
    920    &      $\ell^+ \ell^- \to \nu_e \bar\nu_e H^0 + e^+ e^- H^0$\\
\hline
    960    &      $\ell^+ \ell^- \to Z^0 h^0$\\ 
    970    &      $\ell^+ \ell^- \to Z^0 H^0$\\ 
\hline
    955    &      $\ell^+ \ell^- \to H^+ H^-$\\
    965    &      $\ell^+ \ell^- \to A^0 h^0$\\
    975    &      $\ell^+ \ell^- \to A^0 H^0$\\
\hline
\end{tabular}
\end{center}
\end{table}

    For the  last three, a new subroutine  has been  introduced, \v{HWHIHH},
    whereas the first four make use  of the  implementation of  their SM 
 counterparts, which are based on the subroutines \v{HWHIGW} and \v{HWHIGZ}.  
  
\section{TAUOLA interface}
    An interface to the \TA\ decay package has been added. To use this
    interface, the dummy  subroutines  \v{DEXAY},  \v{INIETC},  \v{INIMAS},
   \v{INIPHX}, \v{INITDK}, \v{PHOINI}, \v{PHOTOS} must be deleted,
   and the parameter \v{TAUDEC}  [\v{'HERWIG'}] must be set to \v{'TAUOLA'}.
   You should then link to both
    \TA\  and  {\small PHOTOS}. The  easiest  way to do  this is  to obtain the
    {\small TAUOLA/PHOTOS} versioning system from 
\small\begin{quote}\tt
            http://wasm.home.cern.ch/wasm/test.html
\end{quote}\normalsize
    and produce a version with the correct  size  of the  \v{HEPEVT}  common 
    block. This system  will provide a  \JS\ demo.  The code for this,
    apart from the \JS\ main program and interface, can then
    be linked to \HW\ (with the above changes) instead of \JS.

    This interface uses the  information from the spin density matrices
    to select the helicity of the decaying $\tau$ leptons if the spin
    correlation algorithm is being used;
    otherwise the  helicity of the decaying $\tau$ is averaged over.

\section{Miscellaneous corrections}
    Corrections have been made affecting the following:
\begin{itemize}
    \item Gaugino pair production in hadron collisions: a bug affecting the
      sign of the centre-of-mass collision angle was corrected.
    
    \item Gaugino/squark   production:  a  bug  in the  matrix  element was
      corrected.

    \item \MS\ Higgs production in hadron-hadron via vector-vector fusion: a 
      bug inducing the use of the mass of the SM Higgs as hard scale  of
      the subprocess as well as its label in the \v{IDHW} array has now been
      taken care of.

    \item Gauge boson pair production:
    \begin{itemize}
      \item A bug in testing the error code after
      the initial state shower has been corrected.

      \item The scale  has been  changed to  the 
      parton-parton  centre-of-mass  energy  from  the  average  of  the 
      produced boson masses, which was used in the previous version.

      \item The W$^+$ and W$^-$ were interchanged in hadronic W$^+$W$^-$
      production. This has been corrected.

    \end{itemize}
      \item Pion Beams. The default MRS nucleon structure functions were  used
      without warning in pion scattering. If you wish  to simulate  pion
      scattering either the older  NSTRU=1,2 pion  sets or PDFLIB should
      be used.
\end{itemize}

\end{document}